\documentclass[aps,preprint,superscriptaddress]{revtex4-1}
\usepackage[T1]{fontenc}
\usepackage[latin9]{inputenc}
\usepackage{mathrsfs}
\usepackage{amsmath}
\usepackage{graphicx}
\usepackage{esint}
\usepackage[unicode=true,pdfusetitle,
 bookmarks=true,bookmarksnumbered=false,bookmarksopen=false,
 breaklinks=false,pdfborder={0 0 1},backref=false,colorlinks=false]
 {hyperref}
\usepackage{breakurl}

\makeatletter

\usepackage{fixmath}

\usepackage{wrapfig}
\usepackage[toc,page]{appendix}
\graphicspath{{images/}}

\usepackage{mathrsfs}

\usepackage{setspace}
\usepackage{breakurl}

\newcommand{\be}{\begin{equation}}
\newcommand{\ee}{\end{equation}}

\makeatother

\begin{document}

\title{ Extreme anisotropy and gyrotropy of surface polaritons in Weyl semimetals}

\author{ Qianfan Chen}

\affiliation{ Department of Physics and Astronomy, Texas A\&M University,
College Station, TX, 77843 USA}

\author{ Maria Erukhimova}

\affiliation{ Institute of Applied Physics, Russian Academy of Sciences, Nizhny Novgorod, 603950, Russia }

\author{ Mikhail Tokman}

\affiliation{ Institute of Applied Physics, Russian Academy of Sciences, Nizhny Novgorod, 603950, Russia }

\author{ Alexey Belyanin}

\affiliation{ Department of Physics and Astronomy, Texas A\&M University,
College Station, TX, 77843 USA}

\begin{abstract}

Weyl semimetals possess unique electrodynamic properties due to a combination of strongly anisotropic and gyrotropic bulk conductivity, surface conductivity, and surface dipole layer. We explore the potential of popular tip-enhanced optical spectroscopy techniques for studies of bulk and surface topological electron states in these materials. Anomalous dispersion, extreme anisotropy, and the optical Hall effect for surface polaritons launched by a nanotip provides information about Weyl node position and separation in the Brillouin zone, the value of the Fermi momentum, and the matrix elements of the optical transitions involving both bulk and surface electron states. 

\end{abstract}

\date{ \today}

\maketitle

A number of recent studies have suggested that Weyl semimetals (WSMs) should have highly unusual optical response originated from unique topological properties of their bulk and surface electron states; see e.g. \cite{kargarian2015, hofmann2016, tabert2016-2, ukhtary2017, felser2017, kotov2018, andolina2018,zyuzin2018, rostami2018, chen2019, narang2019,ma2019,moore2019} and references therein. Their optical response can be used to provide detailed spectroscopic  information about their electronic structure which could be difficult to obtain by any other means. Furthermore, inversion or time reversal symmetry breaking inherent to WSMs makes their optical response strongly anisotropic and/or gyrotropic, enables strong optical nonlinearity, creates anomalous dispersion of normal electromagnetic modes, breaks Lorentz reciprocity, and leads to many other optical phenomena of potential use in new generations of the optoelectronic devices. 

In a recent paper \cite{chen2019}, we investigated general optical properties  of Type I WSMs. Starting from a class of microscopic  Hamiltonians for WSMs with two separated Weyl nodes (\cite{burkov2011,okugawa2014}), we obtained both bulk and surface electron states, derived bulk and surface conductivity tensors, and described the properties of electromagnetic eigenmodes.

Here we focus on one of the most popular and convenient ways to study the properties of novel materials by optical means: a tip-based optical spectroscopy, in which a tip brought in close proximity to the material surface is illuminated with laser light and the linear or nonlinear scattered signal is collected. Strong near-field enhancement at the tip apex may overcompensate the decrease in the volume  of the material where light-matter interaction occurs \cite{raschke2016,raschke2019}. This technique can provide information about surface states and carrier dynamics with about 10 nm spatial and 1 fs time resolution \cite{raschke2019}.  Even more importantly in the context of this paper, nanoscale concentration of the incident light at the tip apex relaxes the optical selection and momentum matching rules. In particular, it allows one to launch various kinds of surface polariton modes which provide valuable information about the properties of both bulk and surface electron states. 

 We use the microscopic model of the optical response of Type I WSMs developed in \cite{chen2019} to predict and describe theoretically the properties of surface polaritons (SPs) launched by a nanotip. We show extreme anisotropy and gyrotropy   in SP radiation pattern originated from Weyl node separation and determined mainly by highly anisotropic surface current and surface dipole layer. We demonstrate anomalous dispersion and extreme sensitivity of SP anisotropy to the frequency of light and Fermi momentum, which makes them a sensitive diagnostics of Fermi arc surface states and may form the basis of efficient light modulators and switches.

\begin{figure}[htb]
\begin{center}
\includegraphics[scale=0.3]{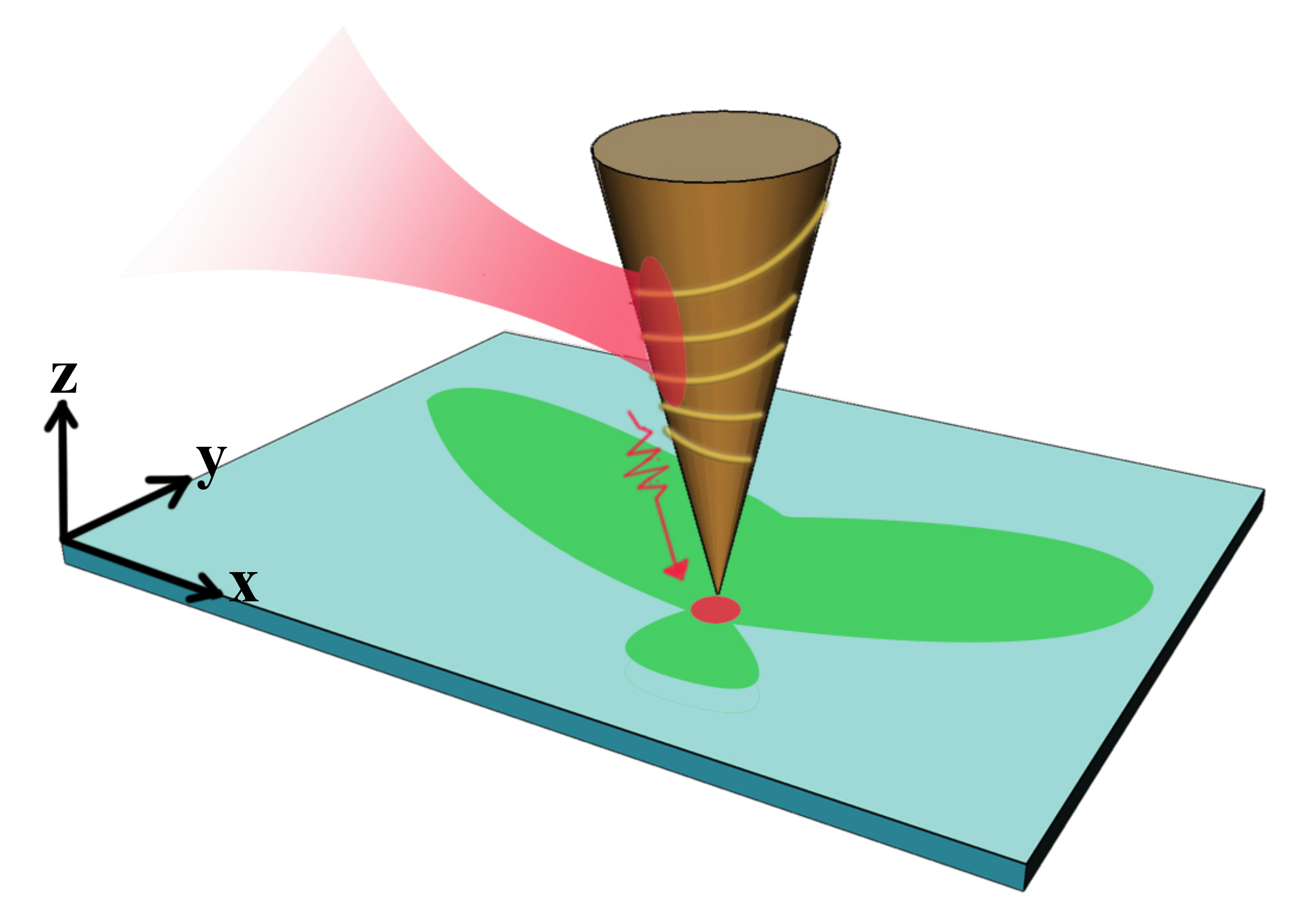}
\caption{A sketch of tip-enabled SP excitation on the WSM surface.  Radiation pattern of SPs is indicated in green for a particular combination of the excitation frequency and Fermi momentum, and for Weyl nodes located along the $k_x$ axis in the Brillouin zone.  }
\label{fig1}
\end{center}
\end{figure}

Figure 1 shows one possible schematic of SP excitation with a gold nanotip. Here the tip apex of $\sim 10$ nm radius is brought to a distance of $\sim 10$ nm from the WSM surface $z = 0$ in order to get access to large SP wavevectors $\sim 10^6$ cm$^{-1}$; see the SP dispersion curves in Fig. 2. A laser beam either illuminates the apex directly (e.g.~\cite{basov}) or excites SPs on the surface of a gold tip via grating, as indicated in the figure \cite{raschke2016,raschke2019}. In the latter case, gold surface plasmon-polaritons  propagate to the apex, experiencing strong adiabatic amplification of the field intensity as they reach the apex \cite{stockman,raschke2016}. Either way, excitation of SPs on a WSM surface is concentrated under the tip within a spot of $\sim 10$ nm. In the linear excitation regime, the frequency spectrum of SPs coincides with the spectrum of an incident laser pulse, whereas the spatial spectrum is extremely broadband, with a cutoff around $10^7$ cm$^{-1}$. The SPs propagate away from the tip, forming a strongly anisotropic radiation pattern which depends on the Weyl node position and separation and the Fermi momentum. They can be detected (converted into an outgoing EM wave) with another tip, a grating, a notch, etc.  

For the most sensitive diagnostics of the electronic structure of WSMs, the frequency of the probing light should be of the order of $\omega \sim v_F b$, where $2b$ is the distance between Weyl nodes in momentum space along $k_x$; see the electron bandstructure plot in Fig.~1 of \cite{chen2019}. In all numerical examples in the paper we assume for definiteness that $ \hbar v_F b = 100$ meV, so the incident laser light should be in the mid-infrared range. However, the formalism presented in the paper is general and does not depend on the choice of incident frequencies as long as the latter are low enough, so that the interband transitions to electron states in remote bands can be neglected. The remote states have a trivial topology and  they are not of interest to this study. 

The Hamiltonian of a WSM with two separated Weyl nodes breaks time-reversal symmetry, which is expected for WSMs with magnetic ordering, e.g. pyrochlore iridates \cite{wan2011}, ferromagnetic spinels \cite{xu2011}, and Heusler compounds \cite{liu2018}. As we showed in \cite{chen2019}, the tensors of both bulk and surface conductivity for Type I WSMs with time-reversal symmetry breaking have a structure corresponding to a biaxial-anisotropic and gyrotropic medium:
\begin{equation}\label{BulkConductivity}
\sigma _{mn}^{B,S}(\omega) = \begin{pmatrix}
\sigma _{xx}^{B,S}  &  0  &  0\\
0  &   \sigma _{yy}^{B,S}  &   \sigma _{yz}^{B,S}\\
0  &   \sigma _{zy}^{B,S}  &   \sigma _{zz}^{B,S}\\
\end{pmatrix}
\end{equation}
where the Weyl points are on the $k_x$ axis,  $
\sigma _{zy}^{B,S}=- \sigma _{yz}^{B,S}$, and superscripts $B$ and $S$ denote bulk and surface conductivity elements, respectively. 
We add background dielectric constant $\varepsilon_b$ due to transitions to remote bulk bands, assuming it to be isotropic and dispersionless at low frequencies, so that the total bulk dielectric tensor is $\varepsilon_{mn}(\omega) = \varepsilon_b \delta_{mn} + 4 \pi i \sigma _{mn}^{B}/\omega$.  The surface conductivity is due to optical transitions between different electron surface states (often called ``Fermi arc states'', although they exist for all momentum states within $k_x^2 + k_y^2 \leq b^2$, not only on the Fermi arc) and between surface and bulk states. It gives rise to the surface current and surface dipole layer. Note a peculiar and most likely unique electrodynamics of WSM surface modes: they are supported by a highly anisotropic and gyrotropic surface current and surface dipole layer sitting on top of a highly anisotropic and gyrotropic bulk WSM material. 
 
 The surface polaritons excited on WSM surfaces parallel to the $x$-axis (assuming that the Weyl points are located along $k_x$)  can be supported by both bulk and surface electron states. However, in the quasielectrostatic approximation $ck \gg \omega$ the SPs are highly localized and the surface states make a dominant contribution to the SP dispersion and radiation pattern \cite{chen2019}. Here $k$ is the magnitude of the SP wavevector in the $z = 0$ plane. Below we outline the calculation of the SP dispersion, energy flux,  and radiation patterns generated by the tip. The detailed derivation is in the Supplemental Material below. 

We model the nanotip-induced excitation source of SPs as an external point dipole, 
\begin{equation}
\mathbf{p}^{e}\left(\mathbf{r},z,t\right)=\mathrm{Re}\left[\mathbf{p}\delta\left(\mathbf{r}\right)\delta\left(z\right)e^{-i\omega t}\right]
\end{equation}
where 
$\mathbf{r}=\left(x,y\right)$. The point source approximation is valid if the tip apex radius and its distance to the surface are smaller than the exponential extent of the excitation field. Our case is borderline as these scales are actually of the same order, but we will still assume a point source for simplicity. One can always generalize the analysis for any spatial distribution of the excitation specific to a given experiment. 
The corresponding external current is 
$ \mathbf{j}_{\omega}^{e}\left(\mathbf{r}\right)=-i\omega\mathbf{p}\delta\left(\mathbf{r}\right)$. 
Within the quasielectrostatic approximation the electric
field of SPs can be defined through the scalar potential: 
$ \mathbf{E}=-\nabla\Phi$, 
where 
\begin{equation}
\Phi(\mathbf{r},z,t)=\mathrm{Re}\left[\Phi_{\omega}\left(\mathbf{r},z\right)e^{-i\omega t}\right].
\end{equation}

Outside the surface, $\Phi_{\omega}$ is described by the Poisson
equation at $z>0$ (in the air or an ambient medium),  
$\nabla^{2}\Phi_{\omega}=0$, 
and Gauss's law in the bulk WSM at $z<0$:
\begin{equation}
\frac{\partial}{\partial x}\left(\varepsilon_{xx}E_{x}\right)+\frac{\partial}{\partial y}\left(\varepsilon_{yy}E_{y}+\varepsilon_{yz}E_{z}\right)+\frac{\partial}{\partial z}\left(\varepsilon_{zz}E_{z}+\varepsilon_{zy}E_{y}\right)=0.
\end{equation}
We assume that the medium above the surface is described by an isotropic dielectric constant $\varepsilon_{up}$. Then, 
the boundary conditions yield 
\begin{equation}
\varepsilon_{up}E_{z}\left(z=+0\right)-D_{z}\left(z=-0\right)=4\pi\rho^{S}=-i\frac{4\pi}{\omega}\left(\frac{\partial}{\partial x}j_{x}^{S}+\frac{\partial}{\partial y}j_{y}^{S}\right)
\end{equation}
where $\rho^{S}$ is the surface charge due to surface electron states and an external source; $j_{x}^{S},$ $j_{y}^{S}$ are the components of the total surface current that are connected with the surface charge 
by the in-plane continuity equation.

The total surface current,
$\mathbf{j}_{\omega}^{S}\left(\mathbf{r}\right)=\mathbf{j}_{\omega}^{l}\left(\mathbf{r}\right)+\mathbf{j}_{\omega}^{e}\left(\mathbf{r}\right)
$, is the sum of the current 
$\mathbf{j}_{\omega}^{l}\left(\mathbf{r}\right)$ 
representing the linear response to $\Phi_{\omega}$ and the current $\mathbf{j}_{\omega}^{e}\left(\mathbf{r}\right)$
induced by the external dipole source. All currents
and charges are on the surface so that we drop the index $S$.

The equations for the scalar potential can be solved by expansion over spatial harmonics in the $(x,y)$ plane: 
\begin{equation}
\mathbf{j}_{\omega}^{e,l}\left(\mathbf{r}\right)=\intop\int\mathbf{j}_{\omega\mathbf{k}}^{e,l}e^{i\mathbf{k\cdot r}}d^{2}k,
\end{equation}
\begin{equation}
\Phi_{\omega}\left(\mathbf{r},z\right)=\intop\int\Phi_{\omega\mathbf{k}}\left(z\right)e^{i\mathbf{k\cdot r}}d^{2}k.
\end{equation}
Here 
$ \mathbf{j}_{\omega\mathbf{k}}^{l}=\hat{\sigma}^{S}\cdot\mathbf{E}_{\omega}\left(z=-0\right)$, $\mathbf{E}_{\omega}\left(z=-0\right)=-i\mathbf{k}\Phi_{\omega\mathbf{k}}(z=-0)$.

A surface dipole layer is formed at the boundary between
the two media. Its dipole moment is oriented along the normal to the surface,
with the space-time Fourier components related to the z-component of the surface current density:
\begin{equation}
d_{z\mathbf{k}}=\frac{i}{\omega}\left[\sigma_{zy}^{S}E_{y}\left(z=-0\right)+\sigma_{zz}^{S}E_{z}\left(z=-0\right)\right].
\end{equation}
The sum of an external and induced dipole creates a jump in the scalar potential $\Phi\left(z\right)$,
\begin{equation}
\Phi_{\omega\mathbf{k}}\left(z=+0\right)-\Phi_{\omega\mathbf{k}}\left(z=-0\right)=4\pi d_{z\mathbf{k}}+\frac{1}{\pi} \mathbf{p}\cdot\mathbf{z_{0}}. 
\end{equation}

The solution for the SP field evanescent in $\pm z$ direction is
\begin{equation}
\Phi_{\omega\mathbf{k}}(z>0)=\phi_{\omega\mathbf{k}}^{up}e^{-\kappa_{up}z},\,\,\,\,\,\,\,\,\Phi_{\omega\mathbf{k}}(z<0)=\phi_{\omega\mathbf{k}}^{W}e^{\kappa_{W}z}.
\end{equation}
Here the spatial harmonics of the potential satisfy algebraic equations 
\begin{equation}
\label{pot1}
\varepsilon_{up}\kappa_{up}\phi_{\omega\mathbf{k}}^{up}+\left[\kappa_{W}\left(\varepsilon_{zz}+\frac{4\pi}{\omega}k_{y}\sigma_{yz}^{S}\right)+ \frac{4\pi \sigma_{yz}^B}{\omega} k_{y}+i\frac{4\pi}{\omega}\left(k_{x}^{2}\sigma_{xx}^{S}+k_{y}^{2}\sigma_{yy}^{S}\right)\right]\phi_{\omega\mathbf{k}}^{W}=\frac{4\pi}{\omega}\mathbf{k}\cdot\mathbf{j}_{\omega\mathbf{k}}^{e}, 
\end{equation}
\begin{equation}
\label{pot2}
\phi_{\omega\mathbf{k}}^{up}+\left(i\frac{4\pi}{\omega}\kappa_{W}\sigma_{zz}^{S}-\frac{4\pi}{\omega}k_{y}\sigma_{zy}^{S}-1\right)\phi_{\omega\mathbf{k}}^{W}=\frac{1}{\pi}\mathbf{p}\cdot\mathbf{z_{0}}
\end{equation}
 and the decay constants $\kappa_{up,W}$ can be found from 
$k^{2}-\kappa_{up}^{2}=0$, $\varepsilon_{xx} k^{2}\cos^2\phi  + \varepsilon_{yy}k^{2} \sin^2\phi-\varepsilon_{zz}\kappa_{W}^{2}=0$, 
where $k_{x}=k\cos\phi$, $k_{y}=k\sin\phi$. 
In the absence of an external dipole, Eqs.~(\ref{pot1},\ref{pot2}) give the dispersion equation for SPs $\mathscr{D}\left(\omega,\phi,k\right)= 0$ (see \cite{supp,chen2019} for an explicit expression). 

The space-dependent expressions for the scalar potential on both sides of the surface are obtained by taking the Fourier transform from $(k, \phi)$ to $(x,y) = (r\cos\theta, r\sin\theta)$. The 2D integrals in momentum space are calculated by series expansion in terms of Bessel functions  and using the integral identity for Bessel functions derived in the Supplemental Material. In the far-field zone of the tip, the scalar potential scales with distance as $\displaystyle \frac{\exp\left[ik_{\omega}(\theta)r \right]}{\sqrt{r}}$. A very cumbersome expression for $k_{\omega}(\theta)$ is derived in the Supplemental Material. Figures 2a,b show the polar plots of the real part of the in-plane SP wavenumber $k_{\omega}(\theta)$ for several values of frequency and Fermi momentum.

\begin{figure}[htb]
\begin{center}
\includegraphics[scale=0.3]{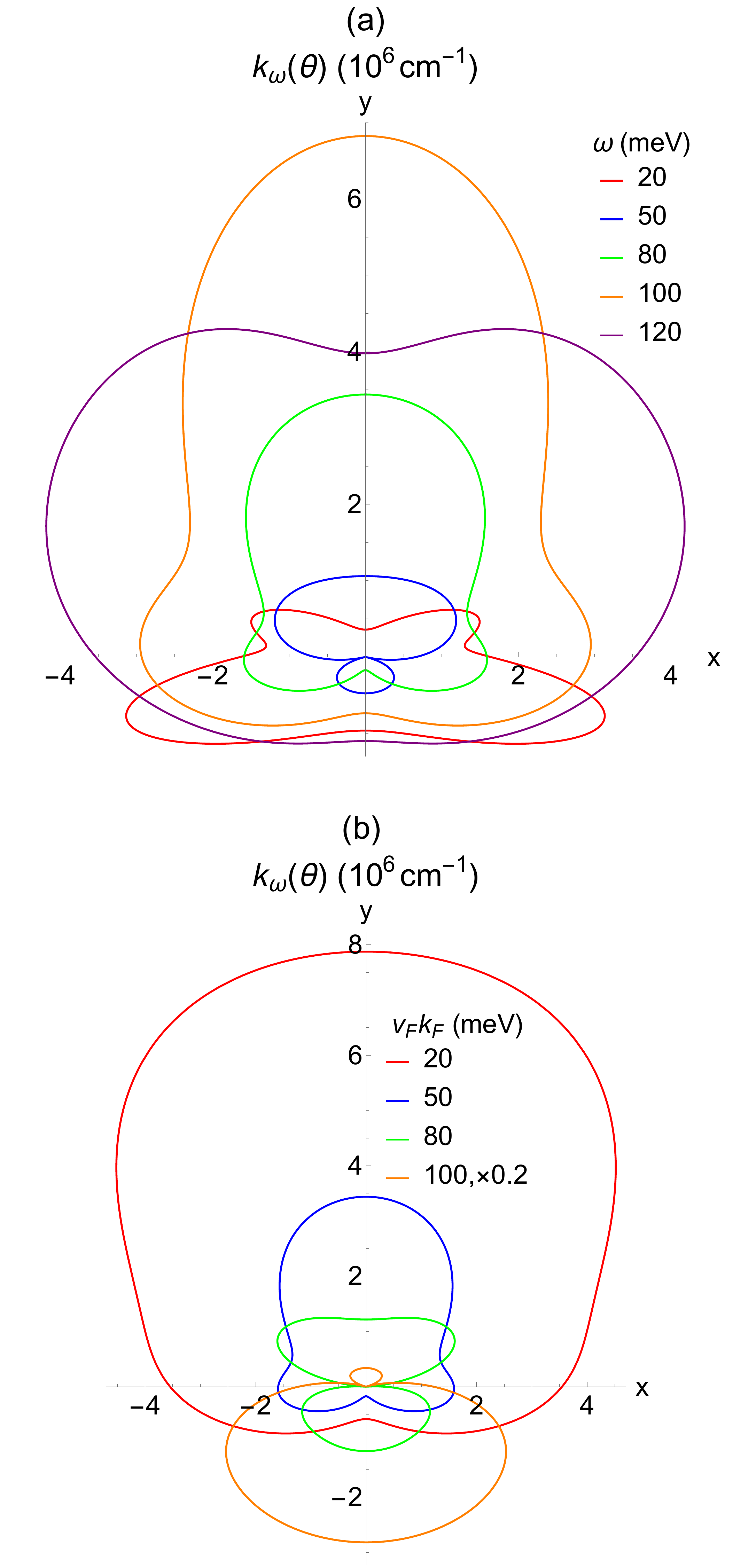}
\caption{Polar plot of the real part of the in-plane SP wavenumber $k_{\omega}(\theta)$ for (a) several values of frequency at a given Fermi momentum $\hbar v_F k_F = 50$ meV and (b) several values of the Fermi momentum at a given frequency $\hbar \omega = 80$ meV.   }
\label{fig2}
\end{center}
\end{figure}

These plots and all plots below were calculated for a vertical dipole orientation. In this case the excitation itself is isotropic in the plane (no $\theta$ dependence) and therefore all anisotropy comes from the properties of topological bulk and surface electron states. The conductivity tensors used in all plots were calculated assuming strongly disordered samples with high SP decay rate $\gamma = 10$ meV. In this case SPs have a low Q-factor: the imaginary part of the wave vector is only a few times lower than the real part. Obviously, in higher quality samples one should expect longer-lived SP excitations with longer propagating lengths, at least at frequencies lower than the Fermi energy-dependent interband transition cutoff determined by the Pauli blocking. We also assumed the background bulk dielectric constant $\varepsilon_b = 10$.

 If the contribution of surface conductivity were ignored and only the bulk carriers were taken into account,  the SPs would have no dispersion at all: their frequency would depend only on the propagation angle but not on the magnitude of the wave vector \cite{chen2019,supp}. Moreover, bulk electron states would support surface EM modes only below the plasma resonance, when the real part of the diagonal components of the bulk dielectric tensor is negative enough. For $\hbar v_F k_F = 50$ meV in Fig.~2a, the plasma resonance is around 50 meV \cite{chen2019}.  SP modes plotted in Fig.~2a show a very strong dispersion in every direction and exist way beyond 50 meV. Therefore they are supported by ``Fermi arc'' surface electron states via surface current sheet and surface dipole that they create in response to the field, with bulk WSM serving mainly as a dielectric substrate. That is why the surface polaritons is a more appropriate term for these surface modes than surface plasmon-polaritons that would exist at  low frequencies below plasma resonance. 
 
 Note strong anisotropy of the wavevector and its extreme sensitivity to the relative values of frequency, Fermi momentum, and Weyl node separation in momentum space. Note also that all plots are symmetric with respect to the $y$-axis, which is perpendicular to the gyrotropy axis $x$. Similar behavior is found in the Poynting flux radiation patterns in Fig.~3. It can be interpreted as the realization of the optical Hall effect. Indeed the symmetry properties of the optical response of the system are determined by the polar symmetry axis vector $\mathbf{a} = \mathbf{n} \times \mathbf{b}$, where the axial gyrotropy vector $\mathbf{b} \parallel \hat{x}_0$ and the polar vector $\mathbf{n} \parallel \hat{z}_0$ is the normal to the surface, so that $\mathbf{a} \parallel \hat{y}_0$. This is in analogy with the Hall effect in which the current direction is determined by the cross product of the axial gyrotropy vector of the magnetic field and the polar vector of the electric field.

To calculate the Poynting flux in a SP wave, we need to go beyond electrostatic approximation. Following the perturbation method detailed in \cite{ryan2018}, we use the Maxwell's equation $\nabla\times\mathbf{B}\left(\omega,\mathbf{r},z\right)=\frac{1}{c}\frac{\partial}{\partial t}\mathbf{D}\left(\omega,\mathbf{r},z\right)$ in each half-space to calculate the magnetic field from the electric field obtained in the electrostatic approximation. 
Then the time-averaged Poynting flux $\mathbf{S}\left(\mathbf{r},z\right)=\mathrm{Re}[\frac{c}{8\pi}\left(\mathbf{E\times B^{\ast}}\right)]$ can be calculated in each half-space. After integrating over $dz$, i.e. $S_{r}\left(r,\theta\right)=\int_{-\infty}^{\infty}S_{r}\left(\mathbf{r},z\right)dz$
we obtain the the time-averaged surface Poynting flux in the far field
zone. The derivation and explicit expression for the Poynting flux can be found in the Supplemental Material.

\begin{figure}[htb]
\begin{center}
\includegraphics[scale=0.35]{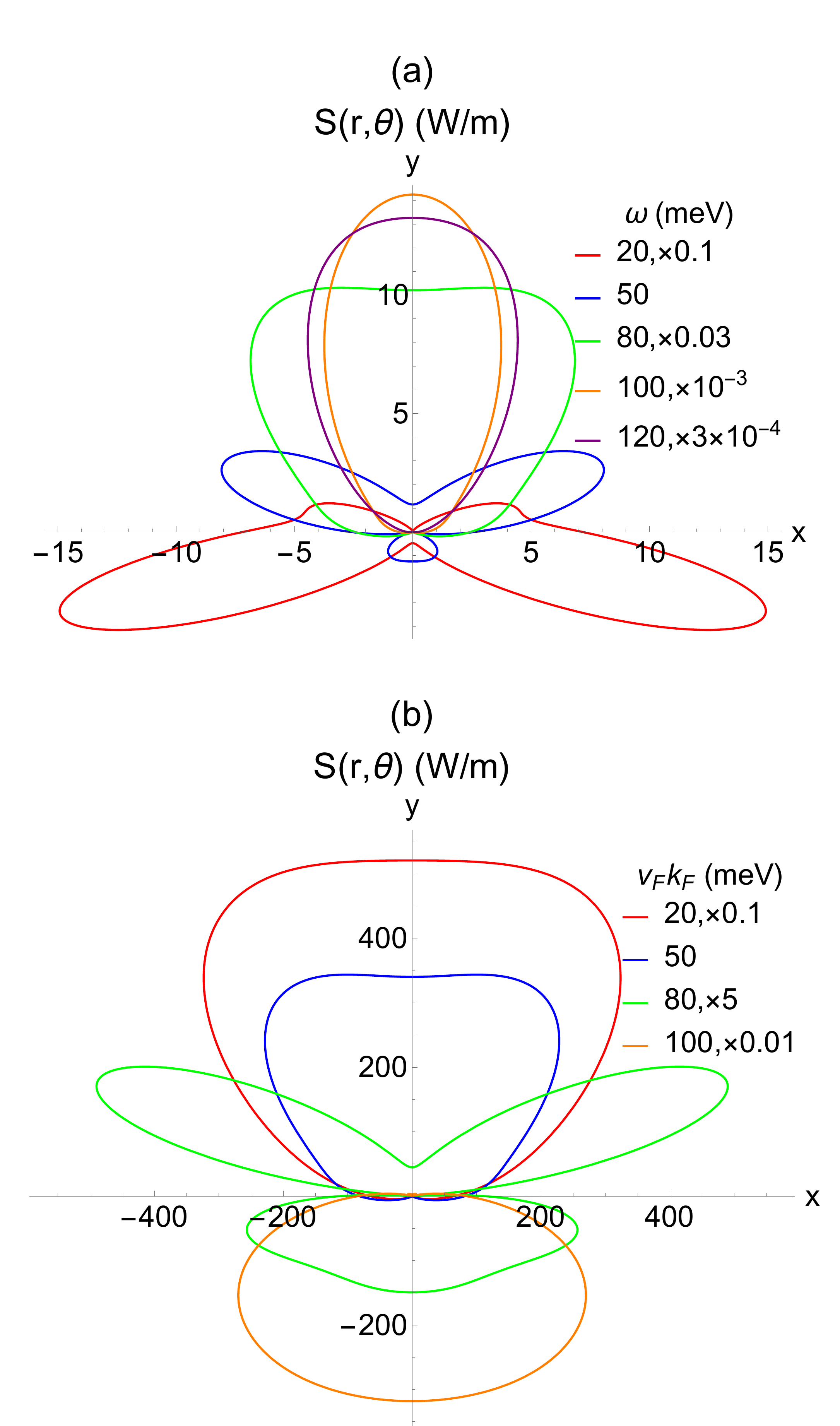}
\caption{Polar plot of the in-plane Poynting vector integrated over the vertical $z$-direction, for (a) several values of frequency at a given Fermi momentum $\hbar v_F k_F = 50$ meV  and (b) several values of the Fermi momentum at a given frequency $\hbar \omega = 80$ meV. The magnitudes of the Poynting flux are multiplied by different numerical factors indicated in the figure, in order to fit to one plot.   }
\label{fig3}
\end{center}
\end{figure}

Figures 3a,b show the radiation pattern of the SPs, namely polar plots of the SP Poynting vector integrated over the vertical $z$-direction, for several values of frequency and Fermi momentum. The numerical values for the SP Poynting flux density in the plots were calculated at a distance of 250 $\mu$m from the tip and assuming that the excitation is created by the pump field of magnitude $10^6$ V/cm localized within (10 nm)$^3$. Such fields are far below damage threshold; for example, in experiments reported in \cite{raschke2019} the pump field under the tip was estimated at $5\times 10^7$ V/cm. Only $1/r$ divergence of the in-plane Poynting vector was included. The actual SP attenuation length is determined by the material quality and is likely to be much shorter than 250 $\mu$m. 

The energy flow of SPs is highly anisotropic and strongly frequency and Fermi momentum-dependent. There is again extreme sensitivity of the radiation pattern to the relative values of frequency, Fermi momentum, and Weyl node separation in momentum space. Furthermore, all plots are symmetric with respect to the $y$ axis and with increasing frequency the SP flux is mainly directed along $\hat{y}_0$. This  is the manifestation of the optical Hall effect induced by Weyl node separation, as discussed above. Note an enhancement in the SP flux at low frequencies in Fig. 3a, related to intraband transitions and Drude-like enhancement of the conductivity, especially its $\sigma_{yy}^S$ element related to free-carrier motion of surface electron states with dispersion $E = \hbar v_F k_y$ \cite{chen2019}. Note also strong enhancement of the Poynting flux at high frequencies around 100 meV due to an increase in the wavenumber $k_{\omega}(\theta)$ and magnitude of the conductivity tensor associated with interband transitions; see the conductivity spectra in \cite{chen2019}. Since the surface states exist only at electron momenta $k_x^2 + k_y^2 < b^2$, at frequencies higher than 200 meV (or for high enough Fermi momenta $k_F > b$) the surface conductivity approaches zero whereas the bulk dielectric tensor approaches its background value. Therefore, there will be no SP modes supported by topological states in this limit $\omega \gg v_F b$, although other kinds of surface polariton modes could still exist due to e.g.~phonon resonances.

In conclusion, we showed that spectroscopy of surface polaritons can be a powerful diagnostics of topological electron states in WSMs. Anomalous dispersion and extreme anisotropy and gyrotropy of SPs launched by a nanotip provides information about Weyl node position and separation, the value of the Fermi momentum, and the matrix elements of the optical transitions involving both bulk and surface electron states. Although the quantitative results in this paper are valid only for magnetic WSMs with time-reversal symmetry breaking, one can still make some qualitative conclusions regarding the optical response of WSMs with inversion symmetry breaking. In particular, one can still expect strong anisotropy of SP propagation, related to the position and orientation of Weyl node pairs in the Brillouin zone. There will be strong dispersion of SPs associated with the presence of Fermi arc surface states. The relative enhancement or suppression of SPs associated with the Fermi edge and interband transitions will be present. The low-frequency response related to massless free carriers will be similar. 

\begin{acknowledgments} 
This work has been supported by the Air Force Office for Scientific Research
through Grant No.~FA9550-17-1-0341. 
M.T. acknowledges the support from RFBR Grant No. 17-02-00387. M.E. acknowledges the support from the Project No. 0035-2018-0006 of the Presidium of the Russian Academy of Sciences. 
\end{acknowledgments}

\section*{Supplemental Material}

On WSM surfaces parallel to the $x$-axis (assuming that the Weyl points are located along $k_x$) the SPPs polaritons can be supported by both bulk and surface electron states. However, in the quasielectrostatic approximation $ck \gg \omega$ the SPPs are highly localized and the surface states make a dominant contribution to the SPP dispersion and radiation pattern \cite{chen2019}. Here $k$ is the magnitude of the SPP wavevector in the $z = 0$ plane. 

We model the nanotip-induced excitation source of SPPs as an external point dipole, 
\begin{equation}
\mathbf{p}^{e}\left(\mathbf{r},z,t\right)=\mathrm{Re}\left[\mathbf{p}\delta\left(\mathbf{r}\right)\delta\left(z\right)e^{-i\omega t}\right]
\end{equation}
where 
$\mathbf{r}=\left(x,y\right)$. 
The corresponding external current is 
$ \mathbf{j}_{\omega}^{e}\left(\mathbf{r}\right)=-i\omega\mathbf{p}\delta\left(\mathbf{r}\right)$. 
Within the quasielectrostatic approximation the electric
field of SPPs can be defined through the scalar potential: 
$ \mathbf{E}=-\nabla\Phi$, 
where 
\begin{equation}
\Phi(\mathbf{r},z,t)=\mathrm{Re}\left[\Phi_{\omega}\left(\mathbf{r},z\right)e^{-i\omega t}\right].
\end{equation}

Outside the surface, $\Phi_{\omega}$ is described by the Poisson
equation at $z>0$ (in the air or an ambient medium): 
\begin{equation} \label{poisson}
\nabla^{2}\Phi_{\omega}=0,
\end{equation}
and Gauss's law inside the WSM at $z<0$:
\begin{equation} 
\label{gauss}
\nabla\cdot\mathbf{D}_{\omega}=0,
\end{equation}
which can be expanded in components as
\begin{equation}
\frac{\partial}{\partial x}\left(\varepsilon_{xx}E_{x}\right)+\frac{\partial}{\partial y}\left(\varepsilon_{yy}E_{y}+\varepsilon_{yz}E_{z}\right)+\frac{\partial}{\partial z}\left(\varepsilon_{zz}E_{z}+\varepsilon_{zy}E_{y}\right)=0.
\end{equation}

We assume that the medium above the surface is described by an isotropic dielectric constant $\varepsilon_{up}$. Then, 
the boundary conditions yield 
\noindent 
\begin{equation}
\varepsilon_{up}E_{z}\left(z=+0\right)-D_{z}\left(z=-0\right)=4\pi\rho^{S}=-i\frac{4\pi}{\omega}\left(\frac{\partial}{\partial x}j_{x}^{S}+\frac{\partial}{\partial y}j_{y}^{S}\right)
\end{equation}
where $\rho^{S}$ is the surface charge due to surface electron states and an external source; $j_{x}^{S},$ $j_{y}^{S}$ are the components of the total surface current that are connected with the surface charge 
by the in-plane continuity equation.

A surface dipole layer is formed at the boundary between
the two media. Its dipole moment is oriented along the normal to the surface,
\[
\mathbf{d}=\mathrm{Re}\left[\mathbf{d}_{\omega}\left(\mathbf{r}\right)e^{-i\omega t}\right],
\]
\[
\mathbf{d}_{\omega}\left(\mathbf{r}\right)=\mathbf{z_{0}}\intop\int d_{z\mathbf{k}}e^{i\mathbf{k\cdot r}}d^{2}k,
\]
where the space-time Fourier components can be related to the z-component of the surface current density,
\begin{equation}
d_{z\mathbf{k}}=\frac{i}{\omega}\left[\sigma_{zy}^{S}E_{y}\left(z=-0\right)+\sigma_{zz}^{S}E_{z}\left(z=-0\right)\right].
\end{equation}
The sum of an external and induced dipole creates a jump in the scalar potential $\Phi\left(z\right)$,
\begin{equation}
\Phi_{\omega\mathbf{k}}\left(z=+0\right)-\Phi_{\omega\mathbf{k}}\left(z=-0\right)=4\pi\left[d_{z\mathbf{k}}+\frac{1}{4\pi^2} \mathbf{p}\cdot\mathbf{z_{0}}\right]. 
\end{equation}

The total surface current,
$\mathbf{j}_{\omega}^{S}\left(\mathbf{r}\right)=\mathbf{j}_{\omega}^{l}\left(\mathbf{r}\right)+\mathbf{j}_{\omega}^{e}\left(\mathbf{r}\right)
$, is the sum of the current 
$\mathbf{j}_{\omega}^{l}\left(\mathbf{r}\right)$ 
representing the linear response to $\Phi_{\omega}$ and the current $\mathbf{j}_{\omega}^{e}\left(\mathbf{r}\right)$
induced by the external dipole source. All currents
and charges are on the surface so that we drop the index $S$.

The equations for the scalar potential can be solved by expansion over spatial harmonics in the $(x,y)$ plane: 
\begin{equation}
\mathbf{j}_{\omega}^{e,l}\left(\mathbf{r}\right)=\intop\int\mathbf{j}_{\omega\mathbf{k}}^{e,l}e^{i\mathbf{k\cdot r}}d^{2}k,
\end{equation}
\begin{equation}
\label{phi}
\Phi_{\omega}\left(\mathbf{r},z\right)=\intop\int\Phi_{\omega\mathbf{k}}\left(z\right)e^{i\mathbf{k\cdot r}}d^{2}k.
\end{equation}
Here 
$ \mathbf{j}_{\omega\mathbf{k}}^{l}=\hat{\sigma}^{S}\cdot\mathbf{E}_{\omega}\left(z=-0\right)$, $\mathbf{E}_{\omega}\left(z=-0\right)=-i\mathbf{k}\Phi_{\omega\mathbf{k}}(z=-0)$.
The inverse transformation is 
\begin{equation}
\mathbf{j}_{\omega\mathbf{k}}^{e,l}=\frac{1}{\left(2\pi\right)^{2}}\intop\int\mathbf{j}_{\omega}^{e,l}\left(\mathbf{r}\right)e^{-i\mathbf{k\cdot r}}d^{2}r.
\end{equation}

The solution for the SPP field evanescent in $\pm z$ direction is
\begin{equation}
\label{phi2}
\Phi_{\omega\mathbf{k}}(z>0)=\phi_{\omega\mathbf{k}}^{up}e^{-\kappa_{up}z},\,\,\,\,\,\,\,\,\Phi_{\omega\mathbf{k}}(z<0)=\phi_{\omega\mathbf{k}}^{W}e^{\kappa_{W}z}.
\end{equation}
Here the spatial harmonics of the potential satisfy algebraic equations 
\begin{equation}
\label{pot1}
\varepsilon_{up}\kappa_{up}\phi_{\omega\mathbf{k}}^{up}+\left[\kappa_{W}\left(\varepsilon_{zz}+\frac{4\pi}{\omega}k_{y}\sigma_{yz}^{S}\right)+gk_{y}+i\frac{4\pi}{\omega}\left(k_{x}^{2}\sigma_{xx}^{S}+k_{y}^{2}\sigma_{yy}^{S}\right)\right]\phi_{\omega\mathbf{k}}^{W}=\frac{4\pi}{\omega}\mathbf{k}\cdot\mathbf{j}_{\omega\mathbf{k}}^{e}, 
\end{equation}
\begin{equation}
\label{pot2}
\phi_{\omega\mathbf{k}}^{up}+\left(i\frac{4\pi}{\omega}\kappa_{W}\sigma_{zz}^{S}-\frac{4\pi}{\omega}k_{y}\sigma_{zy}^{S}-1\right)\phi_{\omega\mathbf{k}}^{W}=\frac{1}{\pi}\mathbf{p}\cdot\mathbf{z_{0}}
\end{equation}
where $g = \frac{4\pi \sigma_{yz}^B}{\omega}$ and the decay constants $\kappa_{up,W}$ can be found from Eqs.~(\ref{poisson}), (\ref{gauss}): 
$k^{2}-\kappa_{up}^{2}=0$, $\varepsilon_{xx} k^{2}\cos^2\phi  + \varepsilon_{yy}k^{2} \sin^2\phi-\varepsilon_{zz}\kappa_{W}^{2}=0$, 
where $k_{x}=k\cos\phi$, $k_{y}=k\sin\phi$. This formalism allows one to add spatial dispersion of the conductivity $\hat{\sigma}^{S}\left(\omega,\mathbf{k}\right)$
and $\hat{\varepsilon}\left(\omega,\mathbf{k}\right)$ if needed,
but we will ignore it below.

In the absence of an external dipole, Eqs.~(\ref{pot1},\ref{pot2}) give the dispersion equation for SPPs  derived in \cite{chen2019}, 
\begin{equation}
\label{disp}
\mathscr{D}\left(\omega,\phi,k\right)=D\left(\omega,\phi\right)-k\Sigma\left(\omega,\phi\right) = 0,
\end{equation}
where 
\begin{equation}
\Sigma\left(\omega,\phi\right) = \frac{4\pi}{\omega}\bigg[\sqrt{\frac{\varepsilon_{xx}\cos^{2}\phi+\varepsilon_{yy}\sin^{2}\phi}{\varepsilon_{zz}}}\left(in_{up}^{2}\sigma_{zz}^{S}-\sigma_{yz}^{S}\sin\phi\right)-n_{up}^{2}\sigma_{yz}^{S}\sin\phi-i\left(\sigma_{xx}^{S}\cos^{2}\phi+\sigma_{yy}^{S}\sin^{2}\phi\right)\bigg],
\end{equation}
and
\begin{equation}
D\left(\omega,\phi\right)=n_{up}^{2}+\varepsilon_{zz}\sqrt{\frac{\varepsilon_{xx}\cos^{2}\phi+\varepsilon_{yy}\sin^{2}\phi}{\varepsilon_{zz}}}+g\sin\phi.
\end{equation}
 
 Note that $\Sigma = 0$ if the surface terms are neglected. Therefore, $D\left(\omega,\phi\right) = 0$ is the dispersion equation of SPPs supported by bulk electron states only. Such modes would have no dispersion since $D\left(\omega,\phi\right)$ does not depend on the SPP wavenumber. Moreover, bulk states would support surface modes only below the plasma resonance, when the real part of the diagonal components of the bulk dielectric tensor is negative enough. For $\hbar v_F k_F = 50$ meV, the plasma resonance is around 50 meV \cite{chen2019}.  SPP modes plotted in Fig.~2 of the main paper show a very strong dispersion in every direction and exist way beyond 50 meV. Therefore they are supported by surface electron states, with bulk WSM serving mainly as a dielectric substrate. 

Including an external source, Eqs.~(\ref{pot1}), (\ref{pot2}) give the Fourier amplitudes of the scalar potential in both half-spaces:
\begin{align}
\label{phi-up}
\phi_{\omega\mathbf{k}}^{up} & =\frac{\frac{4\pi}{\omega}\mathbf{k}\cdot\mathbf{j}_{\omega\mathbf{k}}^{e}\left(\frac{4\pi}{\omega}\sin\phi\sigma_{zy}^{S}+\frac{1}{k}-i\frac{4\pi}{\omega}\sqrt{\frac{\varepsilon_{xx}\cos^{2}\phi+\varepsilon_{yy}\sin^{2}\phi}{\varepsilon_{zz}}}\sigma_{zz}^{S}\right)}{\mathscr{D}\left(\omega,\phi,k\right)}+\frac{1}{\pi}\left(\mathbf{p}\cdot\mathbf{z_{0}}\right)\nonumber \\
 & \times\frac{\varepsilon_{zz}\sqrt{\frac{\varepsilon_{xx}\cos^{2}\phi+\varepsilon_{yy}\sin^{2}\phi}{\varepsilon_{zz}}}+g\sin\phi+\frac{4\pi}{\omega}k\bigg[\sqrt{\frac{\varepsilon_{xx}\cos^{2}\phi+\varepsilon_{yy}\sin^{2}\phi}{\varepsilon_{zz}}}\sigma_{yz}^{S}\sin\phi+i\left(\sigma_{xx}^{S}\cos^{2}\phi+\sigma_{yy}^{S}\sin^{2}\phi\right)\bigg]}{\mathscr{D}\left(\omega,\phi,k\right)},
\end{align}
\begin{equation}
\label{phi-w}
\phi_{\omega\mathbf{k}}^{W}=\frac{\frac{4\pi}{\omega k}\mathbf{k}\cdot\mathbf{j}_{\omega\mathbf{k}}^{e}-\frac{1}{\pi}\left(\mathbf{p}\cdot\mathbf{z_{0}}\right)n_{up}^{2}}{\mathscr{D}\left(\omega,\phi,k\right)}.
\end{equation}
Then the spatial field distributions on both sides of the interface can be obtained from Eqs.~(\ref{phi2}), (\ref{phi-up}), (\ref{phi-w}) by Fourier transform Eq.~(\ref{phi}). 
We will perform integration only in the case of a vertical external Hertz dipole, i.e. $\mathbf{p}=p\mathbf{z_{0}}$, when
$\mathbf{k}\cdot\mathbf{p}=0$ and the source is isotropic in plane of the interface. Therefore, all anisotropy in the SPP propagation comes from the properties of topological electron states. 

The Fourier integral in polar coordinates $(k,\phi)$ in momentum space can be written as 
\begin{align}
\label{phi1} 
\Phi_{\omega}^{\left(+\right)} \equiv \Phi_{\omega}\left(\mathbf{r},z=+0\right) & =\frac{p}{\pi}\intop\int d^{2}ke^{i\mathbf{k\cdot r}}\frac{H\left(\omega,\phi,k\right)}{\mathscr{D}\left(\omega,\phi,k\right)}\nonumber \\
 & \approx -\frac{p}{\pi}\int_{0}^{2\pi}d\phi\frac{1}{\Sigma\left(\omega,\phi\right)}\int_{0}^{\infty}\frac{e^{ikr\cos\left(\phi-\theta\right)}H\left(\omega,\phi,k\right)}{k-k_{\omega}\left(\phi\right)-i\eta_{\omega}\left(\phi\right)}kdk,
 \end{align}
 where $(r,\theta)$ are polar coordinates in real 2D space and we introduced the shortcut notation 
 \begin{equation}
H\left(\omega,\phi,k\right)=D\left(\omega,\phi\right)-n_{up}^{2}+\frac{4\pi}{\omega}k\bigg[\sqrt{\frac{\varepsilon_{xx}\cos^{2}\phi+\varepsilon_{yy}\sin^{2}\phi}{\varepsilon_{zz}}}\sigma_{yz}^{S}\sin\phi+i\left(\sigma_{xx}^{S}\cos^{2}\phi+\sigma_{yy}^{S}\sin^{2}\phi\right)\bigg].
\end{equation}
In the second line of Eq.~(\ref{phi1}) we also introduced the solution to the dispersion equation for SPPs, Eq.~(\ref{disp}) in terms of the  real and imaginary parts of the SPP wave number, $k_{\omega}\left(\phi\right)$ and $\eta_{\omega}\left(\phi\right)$. We will also assume for simplicity that the SPP dissipation is sufficiently weak so that the real part of the solution can be found from 
\begin{equation}
\mathrm{Re}\mathscr{D}\left(\omega,\phi,k_{\omega}\left(\phi\right)\right)\approx 0,
\end{equation}
whereas the imaginary part of the SPP wavenumber can be calculated as 
\begin{equation}
\eta=-\frac{\mathrm{Im}\mathscr{D}\left(\omega,\phi,k_{\omega}\right)}{\left[\frac{\partial\mathrm{Re}\mathscr{D}\left(\omega,\phi,k_{\omega}\right)}{\partial k}\right]_{k=k_{\omega}\left(\phi\right)}}.
\end{equation}
The explicit expression for the SPP wavenumber is 
\begin{equation}
k_{\omega}\left(\phi\right)=\mathrm{Re}\left[\frac{\omega\left(n_{up}^{2}+\varepsilon_{zz}\sqrt{\frac{\varepsilon_{xx}\cos^{2}\phi+\varepsilon_{yy}\sin^{2}\phi}{\varepsilon_{zz}}}+g\sin\phi\right)}{4\pi\left[\sqrt{\frac{\varepsilon_{xx}\cos^{2}\phi+\varepsilon_{yy}\sin^{2}\phi}{\varepsilon_{zz}}}\left(in_{up}^{2}\sigma_{zz}^{S}-\sigma_{yz}^{S}\sin\phi\right)-n_{up}^{2}\sigma_{yz}^{S}\sin\phi-i\left(\sigma_{xx}^{S}\cos^{2}\phi+\sigma_{yy}^{S}\sin^{2}\phi\right)\right]}\right].
\end{equation}
To calculate the integrals in Eq.~(\ref{phi1}), we use the known expansion of the exponent in terms of Bessel functions, 
\begin{equation} 
 e^{iz\cos\alpha} = J_0(z) + 2 \sum_{n=1}^{\infty} i^n J_n(z) \cos(n\alpha),
 \end{equation} 
 which gives 
\begin{align}
\label{phi2}
  \Phi_{\omega}^{\left(+\right)} & =-\frac{p}{\pi}\int_{0}^{2\pi}d\phi\frac{1}{\Sigma\left(\phi\right)}\int_{0}^{\infty}\frac{\left\{ J_{0}\left(kr\right)+2\sum_{n=1}^{\infty}i^{n}J_{n}\left(kr\right)\cos\left[n\left(\phi-\theta\right)\right]\right\} H\left(\omega,\phi,k\right)}{k-k_{\omega}\left(\phi\right)-i\eta\left(\phi\right)}kdk.
  \end{align}
  This integral can be calculated analytically in the far zone of the source dipole, i.e.~at large $kr \gg 1$. In this case the Bessel Functions in Eq.~(\ref{phi2}) oscillate much faster than other $k$-dependent terms in the numerator, so we can take  $H\left(\omega,\phi,k\right)$ out of the integral over $dk$ and replace $k$ with $k_{\omega}(\phi)$ in its argument. After that, the integral over $dk$ can be evaluated using the following integral identity for Bessel functions: 
  \begin{equation}
  \label{iden}
  \int_0^{\infty} \frac{k^nJ_n(kr)}{k^2-k_{\omega}^2 - i0} kdk  = \left(k_{\omega}\right)^n\frac{i\pi}{2} \left( J_n(k_{\omega} r) + iY_n(k_{\omega} r) \right)
  \end{equation}
  Equation (\ref{iden}) can be derived by applying the operator $\left(\frac{1}{r} \frac{d}{dr} \right)^m$ to both sides of the known Hankel transformation \cite{piessens}
   \begin{equation}
  \label{hankel}
  \int_0^{\infty} \frac{J_0(kr)}{k^2-k_{\omega}^2 - i0} kdk  = \frac{i\pi}{2} \left( J_0(k_{\omega} r) + iY_0(k_{\omega} r) \right)
  \end{equation}
and using the recurrent formula 
  \begin{equation}
  \label{rec}
  \left(\frac{1}{z} \frac{d}{dz} \right)^m \left[ z^{-\nu} \mathfrak{G}_{\nu}(z) \right] = (-1)^m z^{-\nu - m} \mathfrak{G}_{\nu+m}(z), 
\end{equation}
where $\mathfrak{G}_{\nu}(z) = J_{\nu}(z), Y_{\nu}(z)$ \cite{abram}. 
Applying Eq.~(\ref{iden}) to the integral over $dk$ in Eq.~(\ref{phi2}) yields 
  \begin{align}
  \Phi_{\omega}^{\left(+\right)} & =-i p\int_{0}^{2\pi}d\phi\frac{k_{\omega}\left(\phi\right)}{\Sigma\left(\phi\right)}H\left[\omega,\phi,k_{\omega}\left(\phi\right)\right]\times\nonumber \\
 & \left\{ \left(J_{0}\left[k_{\omega}\left(\phi\right)r\right]+iY_{0}\left[k_{\omega}\left(\phi\right)r\right]\right)+2\sum_{n=1}^{\infty}i^{n}\cos\left[n\left(\phi-\theta\right)\right]\left(J_{n}\left[k_{\omega}\left(\phi\right)r\right]+iY_{n}\left[k_{\omega}\left(\phi\right)r\right]\right)\right\} \nonumber \\
 & \approx-i p\sqrt{\frac{2}{\pi r}}\int_{0}^{2\pi}e^{ik_{\omega}\left(\phi\right)r}\left\{ e^{-i\frac{\pi}{4}}+2\sum_{n=1}^{\infty}i^{n}e^{-i\left(\frac{n\pi}{2}+\frac{\pi}{4}\right)}\cos\left[n\left(\phi-\theta\right)\right]\right\} \nonumber \\
 & \times\frac{\sqrt{k_{\omega}\left(\phi\right)}}{\Sigma\left(\phi\right)}H\left[\omega,\phi,k_{\omega}\left(\phi\right)\right]d\phi. \nonumber
 \end{align}
 In the last approximate equality we also took an advantage of the fact that in the far zone, namely when the Bessel functions argument $z \gg \left| n^2 - \frac{\pi}{4} \right|$, one can use their asymptotic values \cite{abram} 
 \begin{equation} 
 J_n(z) \approx \sqrt{\frac{2}{\pi z}} \cos\left(z - \frac{n\pi}{2} - \frac{\pi}{4} \right), \; Y_n(z) \approx \sqrt{\frac{2}{\pi z}} \sin\left(z - \frac{n\pi}{2} - \frac{\pi}{4} \right).
 \nonumber 
 \end{equation} 
 Then the integral over $\phi$ can be evaluated by using the delta-function identity: 
 \begin{align} \label{phiplus} 
  \Phi_{\omega}^{\left(+\right)} & =-\frac{p}{\sqrt{\pi}}\sqrt{\frac{2}{r}}\int_{0}^{2\pi}e^{i\left[k_{\omega}\left(\phi\right)r+\frac{\pi}{4}\right]}\frac{\sqrt{k_{\omega}\left(\phi\right)}}{\Sigma\left(\phi\right)}H\left[\omega,\phi,k_{\omega}\left(\phi\right)\right]\sum_{n=-\infty}^{\infty}\cos\left[n\left(\phi-\theta\right)\right]d\phi\nonumber \\
 & =- 2 \sqrt{\pi} p \sqrt{\frac{2}{r}}\int_{0}^{2\pi}e^{i\left[k_{\omega}\left(\phi\right)r+\frac{\pi}{4}\right]}\frac{\sqrt{k_{\omega}\left(\phi\right)}}{\Sigma\left(\phi\right)}H\left[\omega,\phi,k_{\omega}\left(\phi\right)\right]\delta\left(\phi-\theta\right)d\phi\nonumber \\
 & = - \frac{2 \sqrt{\pi} p}{\Sigma\left(\theta\right)}\sqrt{\frac{2k_{\omega}\left(\theta\right)}{r}}H\left[\omega,\theta,k_{\omega}\left(\theta\right)\right]\exp\left[ik_{\omega}\left(\theta\right)r+i\frac{\pi}{4}\right].
\end{align}
Applying the same procedure, we derive the spatial distribution for the scalar potential just below the surface, i.e. inside the WSM:
\begin{equation} \label{phiminus} 
\Phi_{\omega}^{\left(-\right)}\equiv\Phi_{\omega}\left(\mathbf{r},z=-0\right)  =\frac{2\sqrt{\pi} pn_{up}^{2}}{\Sigma\left(\theta\right)}\sqrt{\frac{2k_{\omega}\left(\theta\right)}{r}}\exp\left[ik_{\omega}\left(\theta\right)r+i\frac{\pi}{4}\right].
\end{equation}

To calculate the Poynting flux in a SPP wave, we need to go beyond electrostatic approximation. Following the perturbation method detailed in \cite{ryan2018}, we use the Maxwell's equation $\nabla\times\mathbf{B}\left(\omega,\mathbf{r},z\right)=\frac{1}{c}\frac{\partial}{\partial t}\mathbf{D}\left(\omega,\mathbf{r},z\right)$ in each half-space to calculate the magnetic field from the electric field obtained in the electrostatic approximation:
\begin{equation}
\frac{1}{r}\frac{\partial B_{z}}{\partial\theta}-\frac{\partial B_{\theta}}{\partial z}=i\frac{n_{up}^{2}\omega}{c}\frac{\partial}{\partial r}\Phi_{\omega}^{\left(+\right)}e^{-\kappa_{+}z}\,\,\,\,\,\left(z>0\right),
\end{equation}
\begin{equation}
\frac{1}{r}\frac{\partial B_{z}}{\partial\theta}-\frac{\partial B_{\theta}}{\partial z}=i\frac{\omega}{c}\left[\left(\varepsilon_{xx}\cos^{2}\theta+\varepsilon_{yy}\sin^{2}\theta\right)\frac{\partial}{\partial r}+i\kappa_{-}g\sin\theta\right]\Phi_{\omega}^{\left(-\right)}e^{\kappa_{-}z}\,\,\,\,\,\left(z<0\right),
\end{equation}
where $\kappa_{+}^{2}=k_{\omega}^{2}\left(\theta\right)-n_{up}^{2}\frac{\omega^{2}}{c^{2}}$,
$\kappa_{-}^{2}=$$\frac{\varepsilon_{yy}}{\varepsilon_{zz}}\left[k_{\omega}^{2}\left(\theta\right)-\left(\varepsilon_{zz}-\frac{g^{2}}{\varepsilon_{yy}}\right)\frac{\omega^{2}}{c^{2}}\right].$
In the quasielectrostatic approximation and far field zone, i.e. $c\rightarrow\infty$
and $r\rightarrow\infty$, we have $\kappa_{+}\approx k_{\omega}\left(\theta\right)$,
$\kappa_{-}\approx\left|\mathrm{Re}\left[\sqrt{\frac{\varepsilon_{yy}}{\varepsilon_{zz}}}\right]\right|k_{\omega}\left(\theta\right)$,
$\frac{\partial}{\partial r}\approx ik_{\omega}\left(\theta\right)$. Furthermore,
one can neglect the term $\frac{1}{r}\frac{\partial B_{z}}{\partial\theta}$
in the far field zone. Then we get 
\begin{equation}
B_{\theta}\left(z>0\right)=-\frac{n_{up}^{2}\omega}{c}\Phi_{\omega}^{\left(+\right)}e^{-k_{\omega}\left(\theta\right)z},
\end{equation}
\begin{equation}
B_{\theta}\left(z<0\right)=\frac{\omega}{c}\left(\frac{\varepsilon_{xx}\cos^{2}\theta+\varepsilon_{yy}\sin^{2}\theta}{\left|\mathrm{Re}\left[\sqrt{\frac{\varepsilon_{yy}}{\varepsilon_{zz}}}\right]\right|}+g\sin\theta\right)\Phi_{\omega}^{\left(-\right)}e^{\left|\mathrm{Re}\left[\sqrt{\frac{\varepsilon_{yy}}{\varepsilon_{zz}}}\right]\right|k_{\omega}\left(\theta\right)z},
\end{equation}
\begin{equation}
E_{z}\left(z>0\right)=k_{\omega}\left(\theta\right)\Phi_{\omega}^{\left(+\right)}e^{-k_{\omega}\left(\theta\right)z},
\end{equation}
\begin{equation}
E_{z}\left(z<0\right)=-\left|\mathrm{Re}\left[\sqrt{\frac{\varepsilon_{yy}}{\varepsilon_{zz}}}\right]\right|k_{\omega}\left(\theta\right)\Phi_{\omega}^{\left(-\right)}e^{\left|\mathrm{Re}\left[\sqrt{\frac{\varepsilon_{yy}}{\varepsilon_{zz}}}\right]\right|k_{\omega}\left(\theta\right)z}.
\end{equation}
Therefore the time-averaged Poynting flux $\mathbf{S}\left(\mathbf{r},z\right)=\mathrm{Re}[\frac{c}{8\pi}\left(\mathbf{E\times B^{\ast}}\right)]$
is 
\begin{equation}
S_{r}\left(\mathbf{r},z>0\right)=\frac{n_{up}^{2}\omega}{8\pi}k_{\omega}\left(\theta\right)\left|\Phi_{\omega}^{\left(+\right)}\right|^{2}e^{-2k_{\omega}\left(\theta\right)z},
\end{equation}
\begin{equation}
S_{r}\left(\mathbf{r},z<0\right)=\frac{\omega}{8\pi}k_{\omega}\left(\theta\right)\mathrm{Re}\left[\varepsilon_{xx}^{\ast}\cos^{2}\theta+\varepsilon_{yy}^{\ast}\sin^{2}\theta+g^{\ast}\left|\mathrm{Re}\left[\sqrt{\frac{\varepsilon_{yy}}{\varepsilon_{zz}}}\right]\right|\sin\theta\right]\left|\Phi_{\omega}^{\left(-\right)}\right|^{2}e^{2\left|\mathrm{Re}\left[\sqrt{\frac{\varepsilon_{yy}}{\varepsilon_{zz}}}\right]\right|k_{\omega}\left(\theta\right)z}.
\end{equation}
After integrating over $dz$, i.e. $S_{r}\left(r,\theta\right)=\int_{-\infty}^{\infty}S_{r}\left(\mathbf{r},z\right)dz$
we finally obtain the total in-plane energy flux in the far field
zone: 
\begin{align}
S_{r}\left(r,\theta\right) & =\frac{\omega}{16\pi}\left[n_{up}^{2}\left|\Phi_{\omega}^{\left(+\right)}\right|^{2}+\mathrm{Re}\left(\frac{\varepsilon_{xx}^{\ast}\cos^{2}\theta+\varepsilon_{yy}^{\ast}\sin^{2}\theta}{\left|\mathrm{Re}\left[\sqrt{\frac{\varepsilon_{yy}}{\varepsilon_{zz}}}\right]\right|}+g^{\ast}\sin\theta\right)\left|\Phi_{\omega}^{\left(-\right)}\right|^{2}\right]\nonumber \\
 & =\frac{2\pi^{2}\omega p^{2}n_{up}^{2}k_{\omega}\left(\theta\right)}{\left|\Sigma\left(\theta\right)\right|^{2}r}\left[\left|H\left[\omega,\theta,k_{\omega}\left(\theta\right)\right]\right|^{2}+n_{up}^{2}\mathrm{Re}\left(\frac{\varepsilon_{xx}\cos^{2}\theta+\varepsilon_{yy}\sin^{2}\theta}{\left|\mathrm{Re}\left[\sqrt{\frac{\varepsilon_{yy}}{\varepsilon_{zz}}}\right]\right|}+g\sin\theta\right)\right]
\end{align}

\end{document}